\documentclass{optica-article}
\journal{opticajournal} % for journals or Optica Open
\articletype{Research Article}

\begin{document}
	
	\title{Shot-Noise-Limited Laser Frequency Stabilization Using a High-Resolution Wavelength Meter
	}
	
	%Shot-Noise-Limited Laser Frequency Stabilization Using a High-Resolution Wavelength Meter
	%Sub-10 kHz Frequency Stabilization of a Narrow-Linewidth Laser Using a Precision Wavelength Meter
	\author{
		Mengde Gan,\authormark{1,5} 
		Haoyi Zhang,\authormark{1,5} 
		Xiaodong Tan,\authormark{1,5} 
		Jiaming Li,\authormark{1,2,3,4,*} 
		Le Luo\authormark{1,2,3,4,*}
	}
	
	\address{
		\authormark{1} School of Physics and Astronomy, Sun Yat-sen University, Zhuhai, Guangdong, China 519082\\
		\authormark{2} Guangdong Provincial Key Laboratory of Quantum Metrology and Sensing \& School of Physics and Astronomy, Sun Yat-sen University, Zhuhai, China 519082\\
		\authormark{3} Center of Quantum Information Technology, Shenzhen Research Institute of Sun Yat-sen University, Shenzhen, Guangdong, China 518087\\
		\authormark{4} State Key Laboratory of Optoelectronic Materials and Technologies, Sun Yat-sen University, Guangzhou, China 510275\\
		\authormark{5} These authors contributed equally
	}
	
	\email{\authormark{*}lijiam29@mail.sysu.edu.cn, luole5@mail.sysu.edu.cn}
	
 %% email address is required; see note below about the corresponding author designation
	
	% use {asbstract*} to suppress the copyright line. Copyright information will be added in production
	
	\begin{abstract} 
		We present a compact laser frequency stabilization method by locking a 556 nm laser to a high-precision wavelength meter. Unlike traditional schemes that rely on optical cavities or atomic references, we stabilize the laser frequency via a closed-loop feedback system referenced to a wavelength meter. This configuration effectively suppresses long-term frequency drifts, achieving frequency stability to the wavelength meter’s shot-noise-limited resolution. The system enables sub-hundred-kilohertz stability without complex optical components, making it suitable for compact or field-deployable applications. Our results demonstrate that, with proper feedback design, wavelength meter-based locking can offer a practical and scalable solution for precision optical experiments requiring long-term frequency stability.
	\end{abstract}
	
	%%%%%%%%%%%%%%%%%%%%%%%%%%  body  %%%%%%%%%%%%%%%%%%%%%%%%%%
	\section{Introduction}	
	Laser frequency stabilization is essential for a broad range of applications, including high-resolution spectroscopy\cite{giorgetta2010,ozawa2008,Yoshii:20}, cold atomic and molecular physics\cite{threebody_cooling,langen2024}, quantum information and computing\cite{PengqinxuanLGI, Gong2023}, optical interferometry\cite{Benkler:13,Chekhova2016,Lukin2019}, and optical frequency standards\cite{jiang2011,meiser2009,huntemann2012}. As experimental demands continue to tighten—particularly in ultracold atom physics and quantum metrology—there is a growing need for stabilization systems that combine sub-megahertz precision, long-term reliability, and operational simplicity.
	
	Among the many available stabilization techniques, wavelength meter (WM)-based frequency stabilization has garnered increasing attention for its unique combination of flexibility, compactness, and ease of integration\cite{HuzhongkunHuaZhongKeda, ZhufenAppliedOptics}. Unlike other typical methods that rely on high-finesse optical cavities or atomic/molecular references, WM-based stabilization provides absolute, broadband frequency readout with minimal hardware complexity\cite{CavityLocking, Li2015AppOpt54.3913}. This makes it especially attractive for multi-laser setups, field-deployable systems, and experiments requiring agile reconfiguration or compact footprints. Notably, the advantages of WM-based stabilization have become even more pronounced with the advent of modern fiber and solid-state lasers, which routinely exhibit sub-100 kHz linewidths directly out of the box. In such cases, additional high frequency linewidth narrowing is often unnecessary, shifting the technical emphasis toward stable and accurate frequency referencing. This trend reinforces the relevance and practicality of WM-based stabilization as a cost-effective and scalable solution for contemporary precision applications.
	
	However, despite its appeal, WM-based stabilization faces inherent limitations in frequency stability, particularly at short time scales and under stringent precision requirements\cite{Saleh2015,Couturier2018}. Commercially available systems typically achieve fractional frequency stabilities on the order of $10^{-10}$	over timescales of several seconds to minutes\cite{kim2021,ghadimi2020,utreja2022}. Such performance is insufficient for applications such as laser cooling on narrow intercombination lines or the operation of ultrastable lattice clocks, where absolute frequency fluctuations at the few-kilohertz level are often required\cite{yamaguchi2012,kim2021a}.
	
    The primary factors limiting the performance of WM-based frequency stabilization are  photon shot-noise and slow thermal drifts. In systems based on Fizeau or Michelson interferometers, the wavelength is determined from the spatial position and contrast of interference fringes recorded by a CCD or CMOS sensor\cite{alipieva2001}. Shot-noise introduces statistical fluctuations in the detected signal, resulting in an inherent measurement uncertainty that scales as $\delta f \propto \sqrt{N}$\cite{buttgen2005ccd}, where $N$ is the number of detected photons. This sets a  limit on the short-term frequency precision, even in the absence of other technical noises. Additionally, thermal drifts in the interferometer structure—due to thermal expansion of optical components—lead to slow fluctuations in the measured wavelength, representing a dominant source of long-term instability and accuracy in practical implementations.
	
    In this work, we demonstrate sub-100mkHz-level laser frequency stabilization using a commercially available high-resolution WM in conjunction with a narrow-linewidth 556 nm fiber laser. We begin by conducting a detailed characterization of the noise landscape in our setup, identifying the CCD shot-noise in the WM as the principal constraint on performance. We then implement a suite of system-level optimizations—tuning the interferometer readout parameters, actively suppressing the temperature fluctuation and electronic noise, including 50 Hz power-line interference.
   Under optimized conditions, the system achieves a minimum fractional frequency instability of $1.6 \times 10^{-12}$ at an averaging time of 10,000 s. This represents an improvement of over an order of magnitude compared to the best results previously reported using similar WM-based techniques.\cite{bai2025}.   
   Moreover, the frequency noise spectrum of the stabilized laser remains within a factor of 5 of the WM’s theoretical shot-noise floor across most of the frequency range.   
   These results demonstrate that, with thoughtful noise mitigation and control-loop engineering, WM-based frequency stabilization can closely approach its fundamental noise limit.
   Our results establish a clear performance benchmark for WM-based stabilization and highlight its potential as a compact, robust, and cost-effective solution for precision-demanding applications in quantum optics, atomic physics, and metrology.

	\section{Experimental setup}
	
		\begin{figure}[!ht]
	\begin{center}
		\includegraphics[width=0.8\columnwidth]{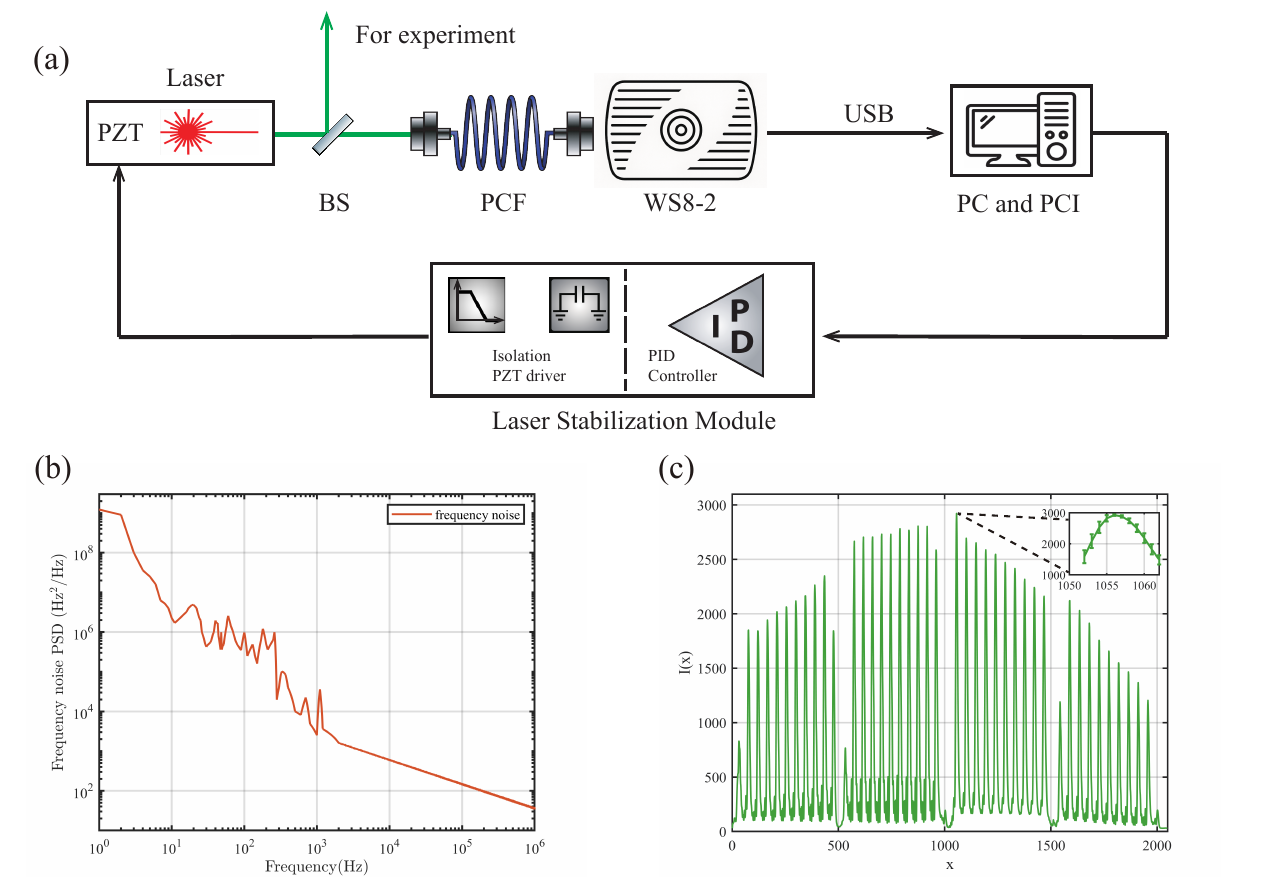}
		\caption{(a) Schematic of the WM-based laser frequency stabilization setup.	The 556 nm laser is stabilized using a WM, PC-controlled digital-to-analog converter (DAC), and PID controller with an isolated low-noise PZT driver. The WM provides real-time frequency readout, converted by the DAC to an analog signal for PID feedback. The isolation driver ensures electrical isolation, noise filtering, and voltage matching for the PZT.
(b) Frequency noise PSD of the free-running 556 nm fiber laser. For Fourier frequencies above 3 Hz, the laser’s intrinsic noise is significantly lower than the shot-noise-limited floor of the WM.
(c) Typical interference fringes recorded by the WM. Inset: close-up of a single fringe peak. Error bars show intensity fluctuations from 10,000 readings.}
		\label{fringe}
	\end{center}
	
\end{figure}
		\begin{figure}[!ht]
			\begin{center}	
				\includegraphics[width=0.8\columnwidth]{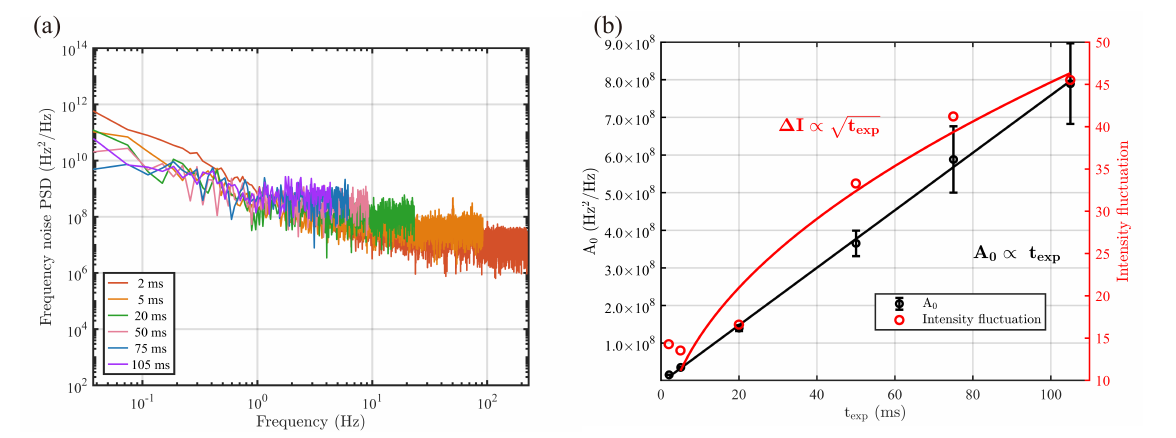}
				\caption{(a) Frequency noise PSD of the WM measured at different CCD exposure times $t_{\mathrm{exp}}$ using the 556 nm fiber laser. 
					(b) Extracted noise floor $A_0$ (black) and measured fringe intensity fluctuations (red) as a function of exposure time $t_{\mathrm{exp}}$.
					$A_0$ increases linearly with $t_{\mathrm{exp}}$, and the intensity fluctuations follow a $\sqrt{t_{\mathrm{exp}}}$ dependence (solid lines), consistent with shot-noise-limited behavior. 
					Error bars indicate standard deviation across 10,000 frames (red) and fitting uncertainty (black).  } 
				\label{expose}
			\end{center}
		\end{figure}
	
The WM-based laser frequency stabilization scheme is illustrated in Fig. \ref{fringe}(a). A 556 nm fiber laser (Model FL-SF-556-0.3-CW, Preciselaser) serves as the light source.  This ytterbium-doped fiber laser, combined with single-pass frequency doubling, delivers high output power while maintaining a narrow linewidth of about 31 kHz. This linewidth is estimated by integrating the frequency noise power spectral density (PSD) over a 1 second window, as shown in Fig. \ref{fringe}(b).
Due to the laser’s built-in feedback loop, high-frequency noise components are effectively suppressed, leaving low-frequency fluctuations as the dominant contribution to the frequency noise. This noise profile makes the laser particularly well-suited for WM-based laser frequency stabilization.
%97 kHz

A commercial WM (Model WS8-2, HighFinesse GmbH) is used as the laser frequency reference. It employs multiple solid-state Fizeau interferometers (FZIs) to perform precise wavelength measurements. The resulting interference fringes are recorded by CCD arrays, generating characteristic fringe patterns [Fig. \ref{fringe}(c)]. The fringe intensity along a transverse axis $x$ is described by
	\begin{equation}
		I(x) = I_0(x)\left[1 +  \cos\left(2\pi x \frac{2\alpha}{\lambda} + \frac{2e}{\lambda}\right)\right].
		\label{fringeeq}
	\end{equation}
Here, $\alpha$ denotes the wedge angle between the reflecting surfaces of the FZI, and $e$ represents the separation thickness at the reference position $x = 0$\cite{Morris1984}. Both $\alpha$ and $e$ are fixed geometrical parameters,  they may exhibit thermal drift. The wavelength $\lambda$ is determined from the fringe pattern. Light is coupled into the wavelength meter via a single-mode photonic crystal fiber, which ensures stable intensity and a well-defined spatial mode. In our setup, the amplitude fluctuation of  $I_0(x)$ remains negligible over 10,000 consecutive measurements, as shown in Fig. \ref{fringe}(c).
From Eq.~\ref{fringeeq}, it is clear that fluctuations in the detected fringe intensity $I(x)$  affect the precision of the extracted wavelength. Two primary noise sources dominate this process: (i) shot-noise-induced intensity fluctuations in the CCD-detected interference fringes, and (ii) low-frequency drifts in the interferometer arising from thermal expansion of optical components. Together, these factors define the resolution limit of WM–based laser frequency stabilization.

To mitigate the impact of shot-noise in $I(x)$, increasing the incident optical power is a straightforward and effective strategy. Higher optical power enhances the signal-to-noise ratio (SNR) in the CCD acquisition, especially when the detector operates near its saturation level. In our system, an incident power of approximately 50 $\mu$W raises the CCD signal to about 90\% of its saturation intensity—well below the manufacturer's damage threshold by more than a factor of 20—thus representing an optimal operating point.
Even under such optimized illumination, photon shot-noise is also increased as $\sqrt{I(x)}$ and imposes a noise floor on the wavelength measurement and hence the achievable frequency resolution. To quantify this limit, we characterize the shot-noise-limited performance of the WM using the narrow-linewidth 556 nm fiber laser. As shown in Fig. \ref{expose}(a), the measured frequency noise PSD confirms that increasing the optical input power shortens the CCD exposure time $t_{\mathrm{exp}}$, thereby suppressing measurement noise and extending the effective measurement bandwidth.
We modeled the frequency noise PSD by fitting the spectrum to
\begin{equation}
	S_{\nu}(f) = A_0 + \frac{B_0}{f^2},
\end{equation}
where $A_0$ represents the white noise floor and $B_0$ captures the random drift ($1/f^2$) noise contribution. As shown in Fig. \ref{expose}(b), the fitted noise floor $A_0$ exhibits a linear dependence on $t_{\mathrm{exp}}$, consistent with the expected shot-noise-limited behavior.

Furthermore, we confirmed the link between CCD shot-noise and wavelength measurement noise by statistically analyzing fringe intensity fluctuations as a function of $t_{\mathrm{exp}}$, as shown in Fig. \ref{expose}(b), the results show that amplitude fluctuations in the interference fringes—driven by CCD shot-noise— contribute to wavelength uncertainty, supporting the shot-noise-limited performance model.

	\begin{figure}[!ht]
	\begin{center}			
		\includegraphics[width=0.8\columnwidth]{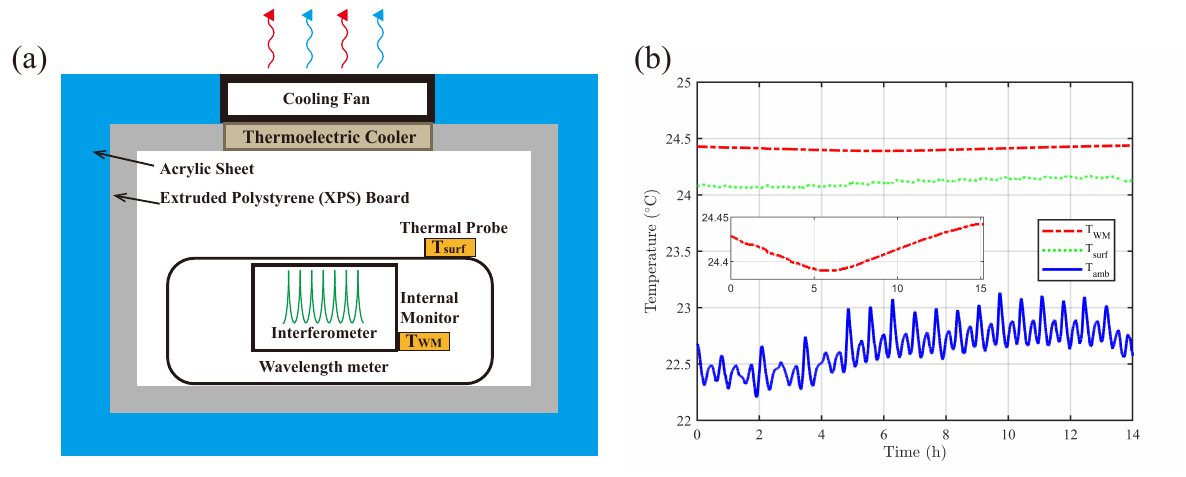}
		\caption{(a) Schematic of the homemade temperature-controlled enclosure for the WM. It consists of an inner acrylic layer (grey), an outer extruded polystyrene shell (blue), and an actively stabilized thermoelectric cooler. A surface-mounted temperature sensor provides feedback for active control. An additional internal sensor monitors the temperature near the FZIs inside the WM.
		(b) Temperature measurements over time. The blue curve shows the ambient temperature $T_{\mathrm{amb}} $, the green curve shows the WM surface temperature $T_{\mathrm{surf}}$, and the red curve shows the internal WM temperature $T_{\mathrm{WM}}$. The inset figure zoom-in the temperature of $T_{\mathrm{WM}}$. } 
		\label{temp}
	\end{center}
\end{figure}

To ensure a thermally stable environment and minimize measurement drift, the WM is housed in a homemade-designed, thermally insulated enclosure, as shown in Fig. \ref{temp}(a). The enclosure incorporates outer panels made of extruded polystyrene for effective thermal insulation\cite{zhang2013} and inner acrylic sheets for mechanical support. Active temperature regulation is implemented using thermoelectric cooler (TEC), with a temperature sensor mounted directly on the WM surface. A PID feedback loop continuously adjusts the TEC output to maintain thermal equilibrium.

Figure \ref{temp}(b) presents the measured temperature stability over a 14-hour period. The green curve indicates that the WM surface temperature, $T_{\mathrm{surf}}$, remains stable within a maximum variation of 0.15°C, despite ambient temperature fluctuations, $T_{\mathrm{amb}} $ (blue curve), exceeding 1 °C. Notably, the WM inner temperature, $T_{\mathrm{WM}}$ (red curve), exhibits even greater stability due to internal thermal isolation, with fluctuations under 0.05°C. This temperature stability corresponds to a maximum wavelength measurement drift of approximately 6 kHz.
This level of thermal stability represents the upper limit achievable through passive insulation. Residual heat generated by internal electronic components—particularly near the FZI—poses a  limitation on further improvement. Nevertheless, the current configuration provides sufficient thermal stability to meet the precision requirements of this study.

Additionally, a particularly challenging noise source in the WM-based laser stabilization setup is the presence of 50 Hz power-line interference, originating from the PC, analog control hardware, and feedback circuitry. This line-frequency noise manifests as a prominent peak in the frequency spectrum, reaching up to $\sim 10^8 \mathrm{Hz}^2/\mathrm{Hz}$ and must be effectively suppressed to ensure stable feedback performance.
To address this, we implemented an electrically isolated analog control module with a low-pass filter featuring a 10 Hz cutoff, as shown in Fig. \ref{fringe}(a). This design effectively isolates the feedback path from external electronic noise and attenuates the 50 Hz component. 
	
\section{Result}
%PID setup, P=xxx, I=xxxx. 
\begin{figure}[!ht]
	\begin{center}			
		\includegraphics[width=0.8\columnwidth]{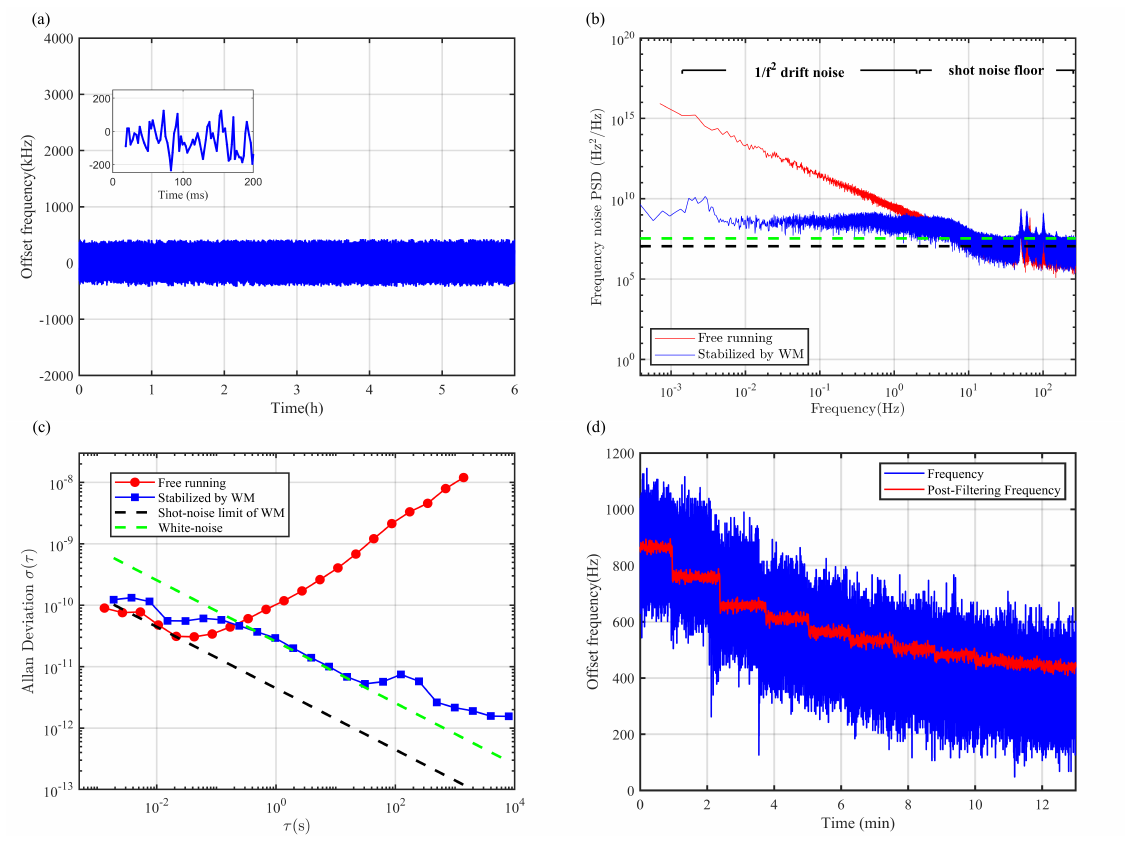}
		\caption{(a) Time trace of the stabilized laser frequency over six hours, demonstrating long-term drift confinement within ±200 kHz. (b) Frequency noise spectrum before (red) and after (blue) stabilization. (c) Allan deviation in the free-running state (red) and after stabilization (blue). Dashed black lines indicate white noise scaling for reference.
		(d) Fine-tuning performance of the stabilized laser frequency. The blue trace shows raw frequency data, while the red trace displays the result after applying a 1 Hz low-pass filter, revealing a resolution improvement to 60 kHz.} 
		\label{allan}
	\end{center}
\end{figure}

Figure \ref{allan} illustrates the performance of the WM-based laser frequency stabilization system. As shown in Fig. \ref{allan}(a), the stabilized laser frequency remains confined within a ±200 kHz window over a continuous six-hour period, demonstrating excellent long-term stability. The standard deviation of the measured frequency is 96.8 kHz.
Figure \ref{allan}(b) presents the frequency noise PSD before and after activating the feedback loop. 
The stabilization system significantly suppresses low-frequency noise, particularly in the sub-hertz range where thermal drifts and mechanical fluctuations dominate. 
Notably, averaging the whole frequency noise PSD after stabilization yields a total noise power of $3.4 \times 10^{7} \mathrm{Hz}^2/\mathrm{Hz}$ (green line), which is only 2.8 times higher than the shot-noise-limited floor of the WM, measured to be $1.2 \times 10^{7} \mathrm{Hz}^2/\mathrm{Hz}$ (black line, averaged above the corner frequency of 25 Hz). This slight excess is primarily attributed to residual electronic noise within the feedback control loop.
Based on the stabilized PSD, the laser linewidth is calculated to be approximately 96.7 kHz, which closely matches the statistical result from Fig. \ref{allan}(a). This consistency confirms that the system operates near the resolution limit imposed by the WM, and that the linewidth broadening—approximately threefold relative to the free-running case—is dominantly limited by the WM’s intrinsic measurement noise.

Figure \ref{allan}(c) compares the Allan deviation $\sigma(\tau)$ of the laser frequency in the free-running (red circles) and stabilized (blue squares) cases. These data are consistent with the noise PSD results in Fig. \ref{allan}(b).  In the free-running case, $\sigma(\tau)$ increases monotonically for $\tau > 10$ ms, reflecting the accumulation of thermal and mechanical drifts from both the WM and the laser. In contrast, the stabilized system exhibits a decreasing trend in $\sigma(\tau)$ up to ten thousand seconds, reaching a minimum of $1.6 \times 10^{-12}$ at $\tau = 10,000$ s. This represents one of the lowest Allan deviations achieved in WM-based laser frequency stabilization to date.
Additional insight can be drawn from Fig. \ref{allan}(c). The black dashed line represents the theoretical shot-noise limit of the WM, following the expected white frequency noise scaling of $\sigma(\tau) = 4.4 \times 10^{-12} /\sqrt{ \tau}$. This behavior is characteristic of photon shot-noise at short averaging times. For $\tau \geq 10^{-2}$ s, the Allan deviation deviates from this ideal limit, as random-walk and drift noise begin to dominate.
The green dashed line represents an elevated white-noise regime after stabilization, described by $\sigma(\tau) = 2.6 \times 10^{-11} /\sqrt{ \tau}$. This additional noise slightly raises the effective noise floor compared to the theoretical limit.

Although the intrinsic noise of the WM limits the ultimate performance of the stabilization system, the frequency tuning resolution of the stabilized laser can still be enhanced by reducing measurement noise through low-pass filtering or signal averaging. Figure \ref{allan}(d) shows the stabilized laser frequency after applying a low-pass filter with a 1 Hz cutoff. The filtered data exhibit significantly reduced fluctuations, demonstrating improved resolution. Visual inspection indicates that the frequency tuning resolution is enhanced to approximately 60 kHz due to the reduced measurement bandwidth. Further improvements are possible by employing filters with even lower cutoff frequencies or more advanced averaging algorithms.

\section{Conclusion and Discussion}
In this work, we have demonstrated a compact and robust laser frequency stabilization system based on a high-resolution WM. By actively locking a narrow-linewidth fiber laser to the WM’s feedback signal—and combining this with excellent thermal shielding, electronic noise suppression, and targeted rejection of 50 Hz line noise—we achieve long-term frequency stability with resolution at the ten-kilohertz level. Measurements of the frequency noise PSD and Allan deviation confirm substantial suppression of both low-frequency drifts and technical noise sources. Notably, the system reaches an Allan deviation minimum of $1.6 \times 10^{-12}$ at 10,000 seconds, representing a roughly tenfold improvement over previously reported WM-based stabilization results. This performance approaches the fundamental shot-noise-limited sensitivity of the WM itself.

From a practical perspective, the WM-based approach offers key advantages: it is compact, hardware-efficient, easy to integrate, supports multi-channel locking, and is broadly tunable in wavelength. The stabilization wavelength can be flexibly adjusted simply by replacing the WM, with no need for major optical reconfiguration.
Looking ahead, several pathways for performance enhancement remain. Upgrading to a WM with higher saturation power and faster readout could further lower the shot-noise limit and enhance locking bandwidth. In addition, better internal thermal management and mechanical layout of the WM device itself could help suppress residual slow drifts.
Given its combination of versatility, simplicity, and high precision, the WM-based stabilization scheme demonstrated here is well suited for a broad range of applications—including laser cooling and trapping, optical tweezer arrays, atom interferometry, and scalable quantum information systems—where reliable, high-resolution laser frequency control is essential.

\section{Acknowledgements}
The authors thanks Jiayu Xiao and Bin Liu for discussion. This work is supported by  NSFC under Grant No.12174458. J.~Li received supports from Fundamental Research Funds for Sun Yat-sen University 2023lgbj0 and 24xkjc015. %L. Luo received supports from Shenzhen Science and Technology Program JCYJ20220818102003006.

\section{Disclosures}
The authors declare no conflicts of interest. 

\bibliography{sample}

\end{document}